# Quasi 1D Structures at the Bi/InAs(100) Surface

Olivier Heckmann,[1, 2] Maria Christine Richter,[1, 2] Jean-Michel Mariot,[3, 4] Laurent Nicolaï,[5] Ivana Vobornik,[6] Weimin Wang,[7] Uros Djukic[1, 2] and Karol Hricovini[1, 2, a)]

[1]*Laboratoire de Physique des Matériaux et des Surfaces, Université de Cergy-Pontoise, 5 mail Gay-Lussac, 95031 Cergy-Pontoise, France*
[2]*DRF, IRAMIS, Service de Physique de l'Etat Condensé, CEA–Saclay, 91191 Gif-sur-Yvette, France*
[3]*Sorbonne Université, CNRS, Laboratoire de Chimie Physique–Matière et Rayonnement, UMR 7614, 4 place Jussieu, 75252 Paris Cedex 05, France*
[4]*Synchrotron SOLEIL, L'Orme des Merisiers, Saint-Aubin, BP 48, 91192 Gif-sur-Yvette, France*
[5]*New Technologies Research Centre, University of West Bohemia, Plzen, Czech Republic*
[6]*Istituto Officina dei Materiali, TASC Laboratory, CNR, 34014 Trieste, Italy*
[7]*Department of Physics, Biology and Chemistry, Linköping University, 581 83 Linköping, Sweden*

[a)]Corresponding author: karol.hricovini@u-cergy.fr

**Abstract.** Thin Bi films are interesting candidates for spintronic applications due to a large spin-orbit splitting that, combined with the loss of inversion symmetry at the surface, results in a band structure that is not spin-degenerate. In recent years, applications for topological insulators based on Bi and Bi alloys have as well attracted much attention. Here we present angle-resolved photoemission spectroscopy studies of the Bi/InAs(100) interface. Bismuth deposition followed by annealing of the surface results in the formation of one full Bi monolayer decorated by Bi nanolines. We found that the building up of the interface does not affect the electronic structure of the substrate. As a consequence of weak interaction, Bi states are placed in the gaps of the electronic structure of InAs(100). We observe a strong resonance of the Bi electronic states close to the Fermi level; its intensity depends on the photon energy and the photon polarization. These states show nearly no dispersion when measured perpendicular to the nanolines, confirming their one-dimensionality.

## I. INTRODUCTION

Bismuth is a group-V semi-metal and is the heaviest non-radioactive chemical element. Therefore Bi is characterized by a very strong spin-orbit coupling, a key ingredient for the Rashba effect, which lifts the spin degeneracy of electronic surface bands, shaping spin-polarized split bands. Thus Bi is an ideal system to study such Rashba split states [1]. In this respect, Bi has been thoroughly studied in its monocrystalline form (see [2] and references therein). Furthermore, Bi is also a parent compound of most topological insulators (TIs) [3], where a strong spin-obit coupling is a vital condition, known up to now. A TI is a non-trivial phase of matter in which the bulk acts like an insulator by opposition to the metallic surface, where the electronic bands are spin polarized. Interestingly, Bi is predicted to be a two-dimensional TI in the form of a single bilayer in the [111] crystallographic orientation [4]. A strong correlation is highlighted with the substrate and it is unclear, to our knowledge, if an independent single Bi bilayer was ever obtained up to now. InAs is an alternative substrate for Bi thin films growth. Here we present angle-resolved photoemission spectroscopy (ARPES) studies of the Bi/InAs(100) interface.

The InAs(100) (4×2)-c(8×2) reconstructed surface exhibits low dimensional self-organized structures. Recently, it has been shown [5] that this reconstruction is formed by one-dimensional (1D) In chains running in the [011] direction. These In chains are separated by about 1.7 nm from each other and are regularly distributed over the whole surface. The region between the chains is occupied by two dimer rows, which are parallel to the In chains [5]. ARPES measurements revealed two surface resonances at normal emission appearing at 31 and 61 eV and their intensity is strongly dependent on photon polarisation.

After deposition on the InAs(100) surface, Bi atoms self-assemble into an ordered pattern of Bi nanolines separated by 4.3 nm on the Bi-stabilized InAs(100) (2×1) structure [6]. Bismuth is reacting only weakly with the In-terminated surface and does not affect the electronic structure of the substrate. In general this is the case for cation-terminated surfaces of III–V semiconductors. As a consequence of weak interaction, Bi electron states are placed in the gaps of the electronic structure of InAs(100).

## II. EXPERIMENT

Photoemission experiments were carried out at the APE beamline of ELETTRA (Trieste, Italy) synchrotron radiation centre. Undoped InAs(100) single crystals were used for all the measurements. First, the In-terminated InAs(100) clean surfaces were prepared after several cycles of Ar-ion bombardment at room temperature and subsequent heating to a temperature of 700 K. The study of the low energy electron diffraction (LEED) patterns showed that we achieve a high-quality In-terminated InAs(100) c(4×2) reconstructed surface. Bismuth was deposited from a Knudsen cell calibrated with a quartz monitor. A photon energy of 31 eV has been used for the valence band measurements.

## III. RESULTS AND DISCUSSION

### Clean InAs(100) Surface

Our measurements of In 4$d$ core levels (not shown) and the clean surface InAs(100) (4×2) are in good agreement with the literature [7]. It should be noted that the samples were cut along the (110) natural cleavage plane from an InAs[100] oriented wafer surface. As the self-organized lines of indium are also in the [110] direction, they are parallel to two sides of the sample. During the photoemission experiments we have selected the linear polarization of photons (electric vector $E$) in the horizontal and vertical directions. In most of the measurements $E$ was parallel to the sample surface. If as well we take into account the measured dispersion of electrons, defined by the hemispherical analyser slit, $\theta_A$, there are four different experimental geometries, as schematically shown in Figs. 1 (a) – (d), upper panels. The rectangles represent the sample with horizontal In lines, the arrow stands for the orientation of the vector $E$, and the thick black lines show the direction of the analyser slit.

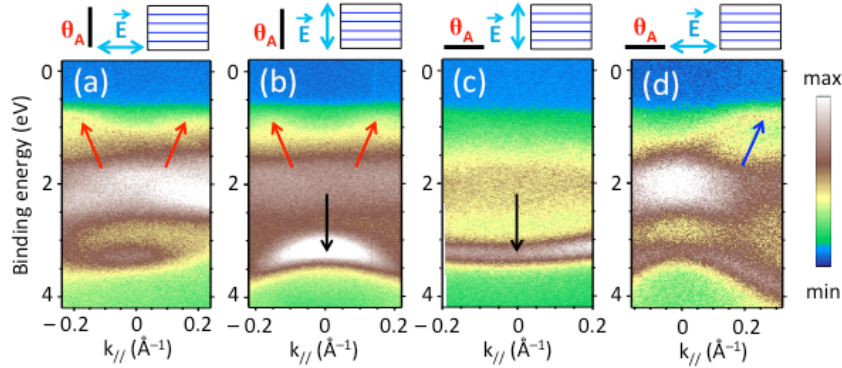

**FIGURE 1.** ARPES spectra of the clean InAs(100) surface for different experimental geometries (a) – (d), given in the upper panels: the thick black line indicates the orientation of the slit in the electron analyser with angular span $\theta_A$; the blue left-right arrows show the polarization of the light (vector $E$); the thin blue lines inside the black rectangles indicate schematically the In nanolines on the InAs(100) surface. The red arrows in (a) and (b), and the blue arrow in (d) show the surface states, whereas the black arrow in (b) and (c) indicates the resonant state.

Figure 1 clearly shows that the relative intensity of the resonant state is highest in the configuration where $E$ is perpendicular to the In lines. These results are in agreement with Ref. [5]. The resonant peak does not disperse with the photon energy, which proves that it is a surface state [5]. The resonance can be attributed to a surface state situated on the edge of the gap of the InAs bulk electronic structure. It follows the dispersion of bulk states and was assigned to back-bonds [8]. The intensity of the resonance falls quickly outside the normal emission when dispersion is measured perpendicular to the In rows [Fig. 1(b)], confirming the 1D character of the In rows. If we rotate the polarization of the light by 90° ($E$ being now parallel to In lines), while keeping the same position of the analyser slit, the intensity of the surface state is completely damped [Fig. 1(a)]. The normal emission spectrum is in excellent agreement with that reported in Ref. [5].

We will now consider the angular dispersion parallel to the In lines with the vector $E$ parallel to the surface. In the case where $E$ is perpendicular to the In lines, we observe, as expected, a high intensity of the resonant state at binding energy of about 3.2 eV [Fig. 1(c)]. Unlike to what we observe for dispersion measured in perpendicular direction to the lines, the intensity of the resonant state does not decrease with $k_{//} \neq 0$. Interestingly, the dispersion exhibits now a convex shape. Rotating the polarization of the light by 90° [Fig. 1(d)], the intensity of the surface state drops and the dispersion becomes concave, probably due to the predominant contribution of bulk states.

We are interested as well by the dispersion of the surface states in the vicinity of the Fermi level (labelled "S1" in Ref. [8]). It is interesting to check whether its dispersion is correlated to the reconstruction of the surface. We remind that the Brillouin zone vector $k_{BZ}$ of the unreconstructed InAs(100) surface is $k_{BZ} = 1.47$ Å$^{-1}$. In Figs. 1(a) and 1(b) one can distinguish a dispersion of a surface state band, indicated by red arrows, with a period of approximately 0.36 Å$^{-1}$, which corresponds very well to 1/4 of the bulk Brillouin zone. In this case, the measured crystallographic direction is [011], in agreement with the multiplicity 4 of the reconstruction. In the direction parallel to the lines In, shown by the blue arrow in Fig. 1(d), the periodicity is twice as large, so tracking the multiplicity 2 of the reconstruction.

## Bi/InAs(100)

In literature, there are only few reports on the Bi/InAs(100) interface. Despite the fact that Bi reacts only weakly with the surfaces of III–V compounds, in particular with InAs, it forms a large variety of reconstructions that depend in a subtle way on the annealing temperature. After deposition of Bi and a subsequent annealing to 520 K, the LEED patterns show a (2×6) reconstruction and scanning tunneling microscopy images provide evidence of the formation of Bi nanolines [6]. The nanolines are separated by 4.3 nm, which represents 10 times the distance of the surface primitive unit cell of InAs(100) (1×1). It is proposed that the nanolines being formed by two parallel chains of Bi contain adjacent Bi dimers, also parallel to the chain. Annealing at higher temperatures causes desorption of bismuth.

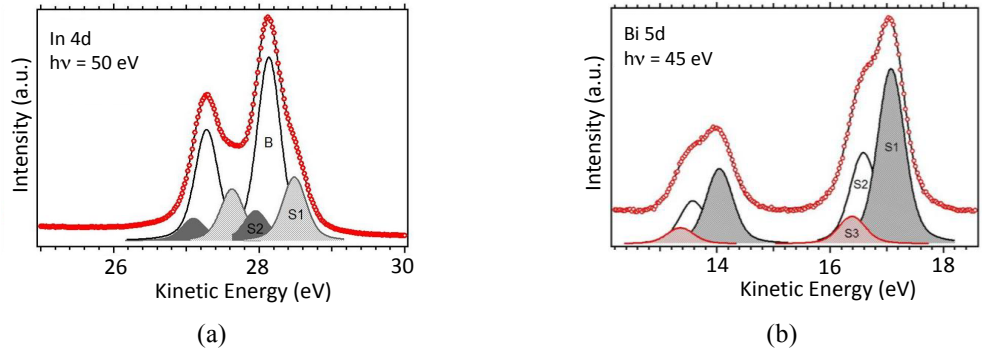

(a)            (b)

**FIGURE 2.** In 4$d$ and Bi 5$d$ core levels for the Bi/InAs(100) interface [panels (a) and (b), respectively]. In the In 4$d$ spectrum, the fitting reveals the bulk (B) component and two surface components (S1 and S2). The components in the Bi 5$d$ spectrum are labelled S1, S2, and S3 as the Bi atoms remain only on the surface.

In 4$d$ and Bi 5$d$ core levels are shown in Fig. 2. The simulation reveals three components for In 4$d$ [Fig. 2(a)], as for the InAs(100) clean surface. B being the bulk component, S1 corresponds to the dimers existing on the InAs clean surface that, as seen in the fit, survive under the Bi layer. The S2 component is attributed to the In atoms connected to the Bi atoms in the second layer. According to theoretical calculations [9], the most stable atomic model for nanoline reconstruction is composed of a Bi monolayer with symmetric Bi–Bi dimers. Therefore, at the origin of component S1 [Fig. 2(b)] are Bi dimers; the S2 component corresponds to the first Bi layer and the S3 component to the second layer placed under the lines of Bi dimers.

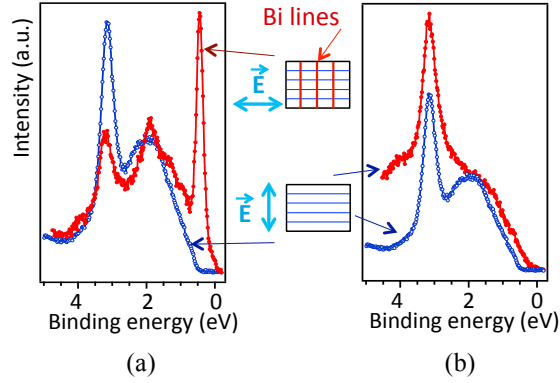

**FIGURE 3.** Normal emission valence band spectra at the resonance photon energy (31 eV) for the clean InAs(100) surface (blue and the Bi/InAs interface (red) measurements for two different orientations of the electric field $E$ of the light with respect to the In lines. The experimental geometry is shown in the panels between the spectra. The thick blue arrow shows the polarization of the light and the thin blue (brown) lines inside the black squares indicate schematically the direction of In (Bi) nanolines. The spectra in (a) were measured for two different relative orientations of the electric field $E$ and the In nanolines as shown by thin blue and brown arrows, whereas both spectra in (b) were measured in the same experimental geometry (bottom panel in the middle).

Figure 3 compares normal emission valence band spectra of the surface with bismuth (red) and of the clean InAs(100) (blue). In both cases, the experimental geometry is optimized to maximize the resonance. Clearly, a strong resonance also exists on the surface with Bi lines. It appears in the spectrum at a different binding energy (0.5 eV) from the one of the clean InAs(100) surface (3 eV). The spectra are normalized to the same intensity of the bulk states, a broad peak around 1.5 eV binding energy, because the bulk states do not have a resonance for the photon energy of 31 eV. The fact that the resonance observed on the clean surface is unchanged even on the surface with bismuth leads to an important conclusion that the lines of indium are not destroyed by the deposition of bismuth.

The finding that the lines of indium are preserved under the bismuth layer is reinforced by the results shown in Fig. 2(b). The spectra for the two surfaces, with bismuth (red) and the InAs(100) surface (blue), are measured in the experimental configuration maximizing the resonant signal of indium lines (the vector $E$ is perpendicular to the In lines). The surface with Bi seems even to amplify the total intensity of the resonant peak. In reality, the shape of the resonant peak is exactly the same. This can be checked superimposing the red and blue peaks. We suggest that the increase in the intensity of the resonant peak is due to secondary electrons. This is quite understandable because the photo-excited electrons from the InAs substrate must cross the Bi layer, which will generate secondary electrons and thus increase the background.

Figure 4(a) shows the dispersions of the clean surface and Fig. 4(b) for the Bi/InAs(100) interface. The spectra at normal emission extracted from the ARPES images are shown in Fig. 4(d). In both cases, the experimental configuration is the same; the field $E$ is perpendicular to the In lines. The spectrum in Fig. 4(a) is the same as in Fig. 1(b), but here in a reduced energy interval; the dispersion of surface states around the binding energy of 1 eV is seen. The white arrow points to the gap between the surface states situated in the vicinity of $k_{//} = 0$. The corresponding normal emission spectrum is shown in green in Fig. 4(c). Figure 4(b) indicates that on the Bi/InAs interface Bi states are situated in the gap of the clean surface, as shown by the black arrow. The normal emission spectrum [in blue in Fig. 4(d)] clearly shows a bump, indicated by a black arrow, in the spectrum and missing on the clean surface (in green).

By rotating the polarization of $E$ by 90° [Fig. 4(c)], we are in a geometry in which the Bi states are in resonance. The normal emission spectrum [in red in Fig. 4(d)] shows an intensity increase of the band at 0.5 eV binding energy.

This band appears as small bumps in the blue curve and is almost invisible in the image in Fig. 4(b). The dispersion measured perpendicular to Bi lines in the resonant geometry [Fig. 4(c)] shows a very small variation in energy at 0.5 eV binding energy. This confirms that the interaction between the Bi lines themselves and with the substrate is very small. This is another confirmation of the 1D character of the Bi deposit.

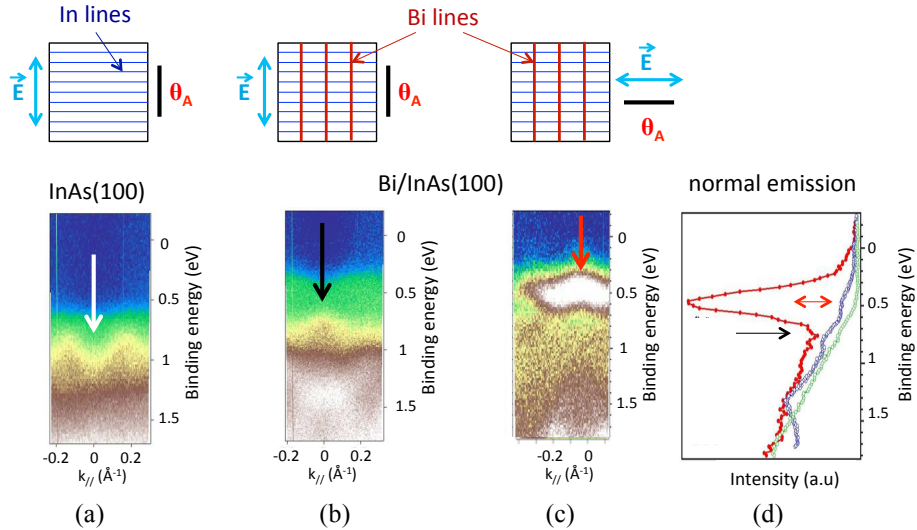

**FIGURE 4.** Upper panels in (a) – (c) show the experimental geometry for the recording of the ARPES spectra for: (a) clean InAs(100) [same as in Fig. 1(b), but here in a reduced energy scale]. The white arrow points to the gap between the surface states; (b) Bi/InAs(100) interface. The black arrow shows the Bi bands that are sitting in the gap shown in (a); (c) Bi/InAs(100) interface measured in a geometry corresponding to the resonance. The red arrow shows strongly enhanced intensity of Bi bands at 0.5 eV binding energy; (d) normal emission spectra extracted from (a): green, (b): blue, and (c): red. Black and red arrows show the same bands as in (b) and (c), respectively.

If, for the configuration in which the field ***E*** and the lines are parallel, we measure the dispersion parallel to lines, the shape of the resonant state changes. As seen in Figs. 1(b) and (c), the dispersion changes from hole-like to electron-like, respectively.

## IV. CONCLUSION

According to our valence band photoemission measurements, it is observed that Bi interacts very weakly with the surface because the original structure of the InAs(100) clean surface remains intact. This is testified by In 4*d* core levels where the surface component S1, attributed to the existing dimers on the clean surface, is surviving under the Bi layer. The study of the valence band reveals the presence of resonant states that are highly sensitive to the photon energy and the polarization of the light, and are in agreement with the quasi-1D structure of the surface. Weak interaction between Bi and the InAs is testified as well by the fact that the Bi electronic states are placed in the gaps of the InAs bands.

## ACKNOWLEDGMENTS

The research leading to these results has received funding from the European Community's Seventh Framework Programme (FP7/2007-2013) under grant agreement n°312284.